\documentclass[%
 reprint,
superscriptaddress,
showpacs,
 amsmath,amssymb,
 aps,
prl,
floatfix,
]{revtex4-1}

\usepackage{graphicx}
\usepackage{dcolumn}
\usepackage{bm}
\usepackage{hyperref}

\usepackage{amsmath}
\usepackage{amsthm}
\usepackage{amsfonts,dsfont}
\usepackage{graphicx}
\usepackage{bbm}
\usepackage{color}
\usepackage{comment}
\usepackage{mathrsfs}

\theoremstyle{remark}

\theoremstyle{remark}

\theoremstyle{definition}

\theoremstyle{lemma}
\newtheorem*{lemma}{Lemma}

\theoremstyle{theorem}
\newtheorem*{proposition}{Proposition}

\theoremstyle{corollary}
\newtheorem*{corollary}{Corollary}

\theoremstyle{conjecture}

\newcommand{\bra}[1]{\langle #1 |} 
\newcommand{\ket}[1]{| #1 \rangle } 
\newcommand{\tr}[0]{\mathrm{tr}} 
\newcommand{\upd}[0]{\mathrm{d}}
\newcommand{\expect}{\mathbb{E}}

\definecolor{cbl}{rgb}{0,0,1}

\definecolor{crd}{rgb}{1,0,0}
 
\begin{document}


\title{Zooming in on Quantum Trajectories}

\author{Michel Bauer}
\email{michel.bauer@cea.fr}
\affiliation{
Institut de Physique Th\'eorique, CEA Saclay and CNRS, Gif-sur-Yvette, France
}
\author{Denis Bernard}
\email{denis.bernard@ens.fr}
\author{Antoine Tilloy}
\email{antoine.tilloy@ens.fr}
\affiliation{Laboratoire de Physique Th\'eorique de l'ENS,
CNRS and Ecole Normale Sup\'erieure de Paris, France
}

\date{\today}

\begin{abstract}
We propose to use the effect of measurements instead of their number to study the time evolution of quantum systems under monitoring. This time redefinition acts like a microscope which blows up the inner details of seemingly instantaneous transitions like quantum jumps. In the simple example of a continuously monitored qubit coupled to a heat bath, we show that this procedure provides well defined and simple evolution equations in an otherwise singular strong monitoring limit. We show that there exists anomalous observable localised on sharp transitions which can only be resolved with our new effective time. We apply our simplified description to study the competition between information extraction and dissipation in the evolution of the linear entropy. Finally, we show that the evolution of the new time as a function of the real time is closely related to a stable L\'evy process of index $1/2$.
\end{abstract}

\pacs{03.65.Ta, 03.65.Yz, 05.40.-a} 

\maketitle

\paragraph{Introduction ---}
Quantum monitoring equations play a key role in modern theoretical quantum physics and are used widely in control \cite{wiseman1993,wiseman1994,doherty1999}, quantum information \cite{jacobs2003,combes2008,combes2015}, and even foundations \cite{bassi2003,bassi2013}. They describe a system subjected to iterated or continuous measurements and can be used to treat a large variety of experimental setups e.g. in cavity QED \cite{guerlin2007,gleyzes2007} and circuit QED \cite{murch2013,weber2014}. 

An interesting regime which has been explored recently \cite{jumps} is that of ``tight'' monitoring, i.e. the limit when the measurement strength (or frequency) dominates the evolution. This regime is characterised by the emergence of quantum jumps similar to what could be seen in early monitoring experiments \cite{nagourney,bergquist,sauter} but with a richer and subtler structure in the fluctuations \cite{spikes}.
What makes this limit interesting is that it is expected to yield a finer description of Von Neuman measurements and quantum jumps. It is however difficult to study because the evolution equations become singular and ill-defined with infinitely sharp transitions when the monitoring tightness goes to infinity. In this article, we propose a time redefinition which allows to take the latter limit exactly at the evolution level without losing any information.

Although the rest of the article deals with an example of continuous quantum trajectory, let us introduce our idea in a discrete setting for simplicity. In the discrete case, quantum monitoring is simply a succession of (generalised) discrete measurements and evolution. After each measurement, the density matrix is updated conditionally on the result. The sequence of the system density matrices $\{\rho_n\}$ after each measurement is a discrete quantum trajectory. If we suppose that the measurements are carried out regularly, the natural time to parametrise the evolution is simply proportional to the number of measurements $n$. Here, we propose a new parametrisation --different from the real physical time-- proportional to the \emph{effects} of the measurements on the system:
\begin{equation}\label{eq:discrete}
t_n:=\sum_{1 \leq m \leq n} \text{Tr}\,[ (\rho_m-\rho_{m-1})^2],
\end{equation}
which is simply the quadratic variation of the density matrix \footnote{Other prescriptions with the same quadratic scaling are possible, e.g. $\Delta t_n= \text{Tr}\,[ (D(\rho_{n})-D(\rho_{n-1}))^2]$ where $D(\cdot)$ denotes the diagonal part of a matrix in the measurement pointer basis.}. Because this new effective time will flow more when the system evolves abruptly, it will resolve the inner structure of sharp transitions. Notice that as $\rho_n$ is a function of the measurement results, $t_n$ is a quantity which can be computed from standard experimental data.

\begin{figure}
\centering
\includegraphics[width=0.99\columnwidth]{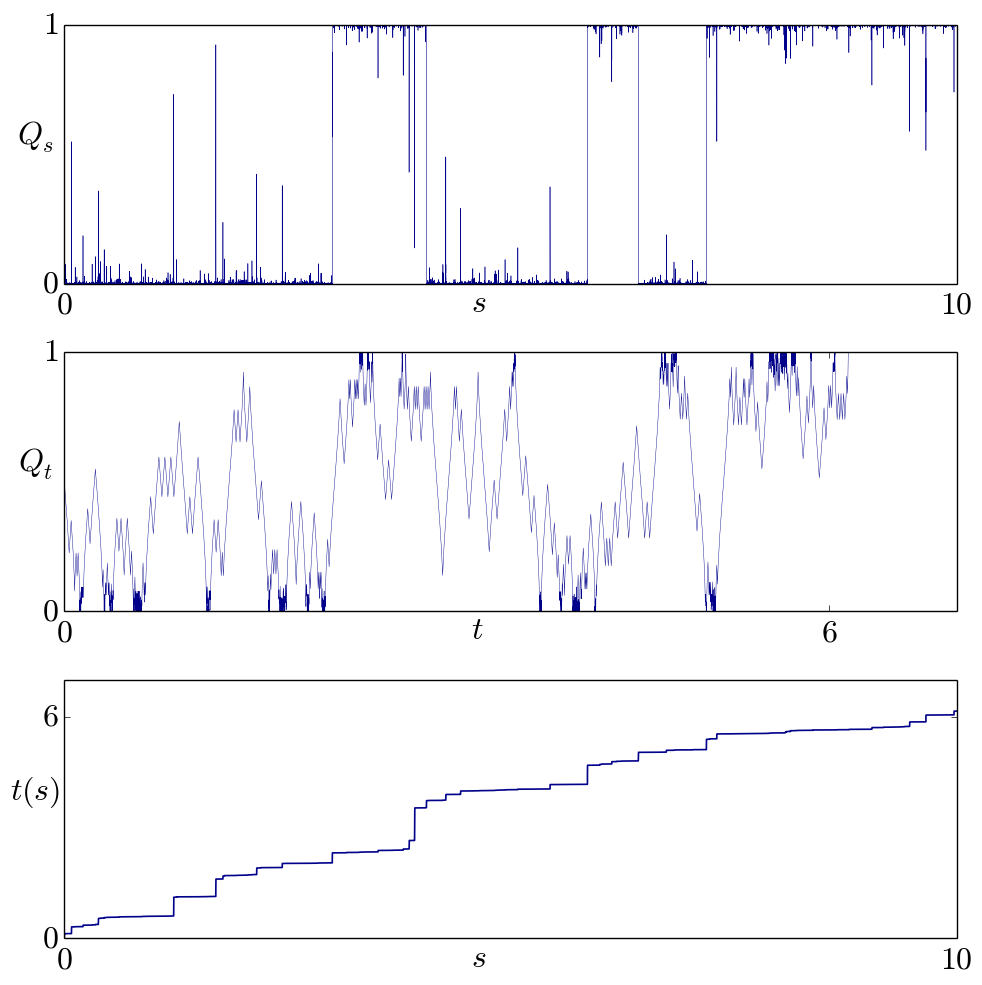}
\caption{\textit{
Discrete quantum trajectories in real and effective time.
\emph{Top}: The evolution of the ground state probability $Q$ in real time $s$ shows sharp jumps and spikes. \emph{Center}: The evolution of $Q$ in effective time $t$ allows to resolve what happens outside the boundaries $0$ and $1$, the details are unfolded. \emph{Bottom}: Effective time as a function of the real time. The plots are shown for the same realisation with $\varepsilon=0.3$, $\Delta s=10^{-5}$, $p=0.5$ and $\lambda=1$.
}}
\label{fig:plot0}
\end{figure}

Let us be more concrete and specify the procedure in a simple example of evolution with discrete measurements. We consider a two-level system coupled to a thermal bath with a density matrix $\rho_s$ obeying:
\begin{equation}
\partial_s \rho_s =\mathcal{L}(\rho_s)
\end{equation}
where $\mathcal{L}$ is a Lindblad operator of the form:
\begin{equation}
\begin{split}
\mathcal{L}(\rho) = &\lambda p \Big(\sigma_- \rho \sigma_+ -\frac{1}{2}\{\sigma_+\sigma_-,\rho\}\Big)\\
&+\lambda (1-p) \Big(\sigma_+ \rho \sigma_- -\frac{1}{2}\{\sigma_-\sigma_+,\rho\}\Big)
\end{split}
\end{equation}
$\lambda$ is interpreted as the thermal relaxation rate and $p$ the average population of the ground state at equilibrium.
Every $\Delta s$, the system energy is weakly measured, i.e. it is subjected to the discrete random map:
\begin{equation}
\rho_{n\Delta s^+} = \frac{ B_\pm \rho_{n\Delta s^-} B^\dagger_\pm}{\text{Tr}\left[B_\pm \rho_{n\Delta s^-} B^\dagger_\pm\right]} 
\end{equation}
with probability $\text{Tr}\left[B_\pm \rho_{n\Delta s^-} B^\dagger_\pm\right]$
where 
\begin{equation}
B_{\pm}=\left(\begin{array}{cc}
 \sqrt{1\pm\varepsilon}& 0  \\
0 & \sqrt{1\mp \varepsilon}
\end{array}\right)
\end{equation}
with $\varepsilon \in ]0,1[$ coding for the measurement strength and $B_+^\dagger B_+ + B_-^\dagger B_-=\mathds{1}$. For a fixed value of $\varepsilon \neq 1$, when the real time $\Delta s$ between two measurements goes to zero, the fast weak measurements should behave like a strong measurement and it is this limit we are interested in. Numerical simulations for $Q=\bra{0}\rho\ket{0}$ are shown in Fig. \ref{fig:plot0}.
They shows --at least visually-- that the prescription of eq. \eqref{eq:discrete} indeed allows to blow up the details of the sharp fluctuations. In the continuous setting, the reparametrisation in effective time will have the extra advantage of yielding simpler equations which can be analysed in detail.
\paragraph{Model ---}

In this article, we will focus on one of the simplest instances of continuous quantum trajectory equations \cite{barchielli1986,belavkin1992,jacobs2006,wiseman2009} which contains the quintessential subtlety of the fast measurement limit while being analytically manageable (see e.g. \cite{lmp}). It describes the continuous energy monitoring of a two-level system coupled in the same way as before to a thermal bath and reads:
\begin{equation}\label{eq:realtime}
\upd Q_s= \lambda (p-Q_s) \, \upd s + \sqrt{\gamma} Q_s (1-Q_s) \, \upd W_s,
\end{equation}
where again $Q=\bra{0}\rho\ket{0}$ is the probability to be in the ground state, $s$ the real physical time and $\gamma$ codes for the measurement rate, i.e. the rate at which information is extracted from the system. As in the discrete case, $\lambda$ is the thermal relaxation rate and $p$ the average population of the ground state at equilibrium. The stochastic process $W_s$ is a Brownian motion which echoes the intrinsic quantum randomness of continuous measurements. Notice that we do not consider the non-diagonal coefficients of the density matrix because they have no effect on the probabilities in this model and are anyway exponentially suppressed. Our objective is to see what the limit of equation \eqref{eq:realtime} is when $\gamma\rightarrow +\infty$, i.e. when the measurement strength becomes infinite.

\begin{figure}
\centering
\includegraphics[width=0.99\columnwidth]{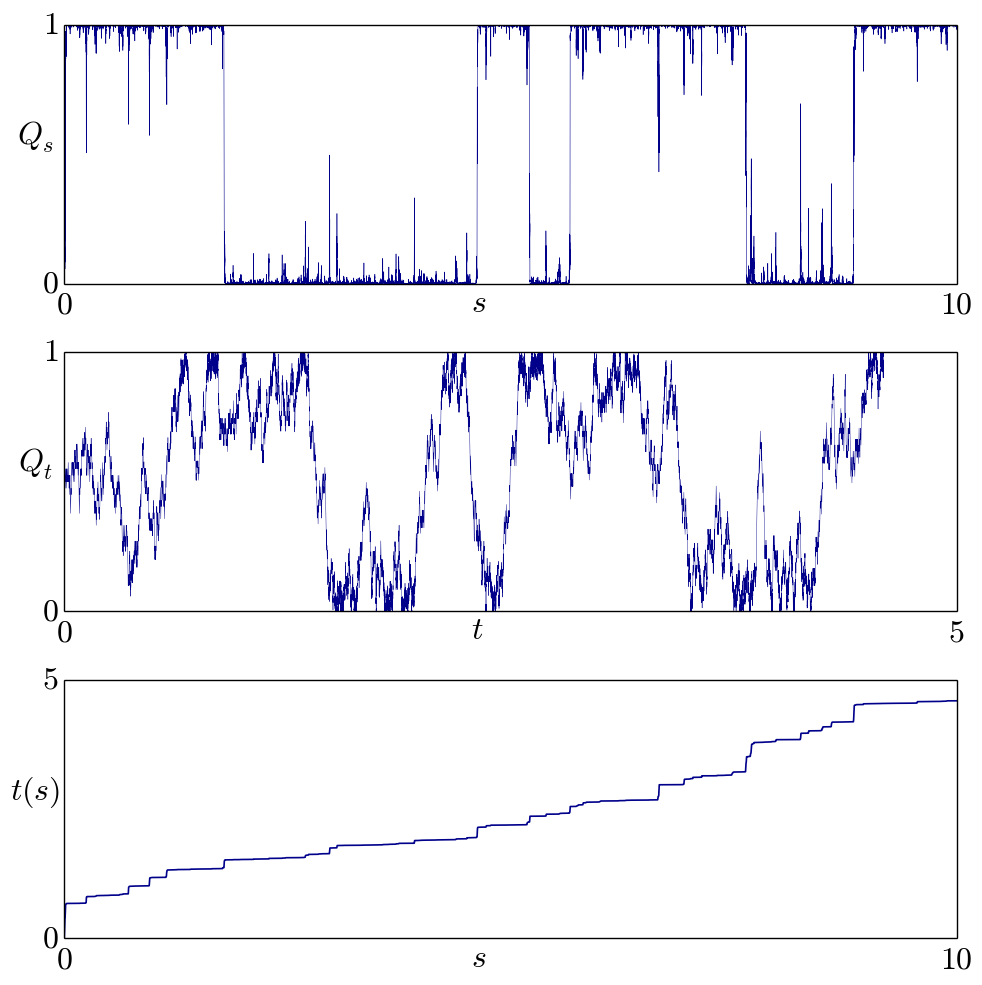}
\caption{\textit{
Continuous quantum trajectories in real and effective time.
\emph{Top}: The evolution of the ground state probability $Q$ in real time $s$ shows sharp jumps and spikes. \emph{Center}: The evolution of $Q$ in effective time $t$ looks like a reflected Brownian motion without sharp transitions. \emph{Bottom}: Effective time as a function of the real time. The plots are shown for the same realisation with $\gamma=200$ (which looks like $\gamma\rightarrow +\infty$), $p=0.5$ and $\lambda=1$.
}}
\label{fig:plot1}
\end{figure}

The trajectories of equation \eqref{eq:realtime} become very singular when $\gamma \rightarrow +\infty$ with sharp jumps between plateaus decorated with instantaneous excursions dubbed \emph{spikes} (see Fig. \ref{fig:plot1}) \footnote{One can give a heuristic argument for the existence of spikes. Near a boundary say $Q=0$, $\upd Q_s\simeq \lambda p \upd s +\sqrt{\gamma}Q_s \upd W_s$. One can show that for large $\gamma$, this means that the distribution of $Q$ is given by $\mathds{P}[Q<q]=e^{-2/(\gamma\lambda p q)}$ and that $Q_{s+1/\gamma}$ is weakly correlated with $Q_s$. As a result, the maximum $M_{a,b}$ of $Q$ in an interval $[a,b]$  has a distribution which can be approximated by $\mathds{P}[M_{a,b}<m]\simeq\left(e^{-2/(\gamma\lambda p m)}\right)^{\gamma (b-a)}$. The $\gamma$ cancel out which means that excursions with height of order $0$ (but vanishing width) persist in the limit: these are the spikes.}. Although the stochastic differential equation (SDE) \eqref{eq:realtime} in real time has no well defined limit when $\gamma \rightarrow +\infty$, the plot of the process heuristically has one in the sense that the extrema of the spikes are a sample of a $\gamma$ independent Cox process. This limiting process can be studied directly as was done in \cite{spikes} but the analysis with the effective time provides a cleaner derivation in addition with the discovery of an even finer anomalous \footnote{We borrow this terminology from quantum or statistical field theory and from fluid turbulence. For instance, observables linked to dissipative processes in Burgers' turbulence are localised on velocity shocks \cite{polyakov1993,polyakov1995,bernard1998}. In a way similar to Burgers' turbulence, we can consider ``anomalous observables'' localised on spikes and jumps.} structure.

\paragraph{Results ---} The effective time we are going to use to redefine the process is:
\begin{equation}
t(s):=\int_0^s (\upd Q_u)^2= \gamma \int_0^s Q_u^2(1-Q_u)^2 \upd u,
\end{equation}
which is the continuous analog of the prescription \eqref{eq:discrete}.
With this effective time $t$, equation \eqref{eq:realtime} becomes:
\begin{equation} \label{eq:thermal-t}
\upd Q_t=\frac{\lambda (p-Q_t) }{\gamma \, Q_t^2\,(1-Q_t)^2} \, \upd t + \upd B_t,
\end{equation}   
where $B_t$ is a Brownian motion (as a function of $t$) related to $W_s$ by $\upd B_t:=\sqrt{\gamma}\, Q_s(1-Q_s) \, \upd W_s$. The crucial feature of the new evolution equation is that, for large $\gamma$, the first term is negligible as long as $Q_t$ is not very close to $0$ or $1$. It is positive (resp. negative) when $Q_t$ is close to $0$ (resp. close to $1$). Intuitively, this term will survive only as a boundary condition preventing $Q$ from crossing $0$ and $1$ and $Q_t$ will be a simple Brownian motion in the bulk $]0,1[$. 

\begin{proposition}
When $\gamma\rightarrow+\infty$:
\begin{itemize}
\item[(i)] $Q_t$ is a Brownian motion reflected at $0$ and $1$.
\item[(ii)] The linear time $s$ can be expressed as a function of the effective time $t$:
\begin{equation}\label{eq:st}
s(t) = \frac{L_t}{\lambda p}+ \frac{U_t}{\lambda (1-p)}
\end{equation}
where $L_t$ and $U_t$ are the \emph{local times} spent by $Q_t$ respectively in $0$ and $1$.
\end{itemize} 
\end{proposition}
For a Brownian like process $X_t$, the local time $L_t$ at $0$ is defined informally by $L_t :=\int_0^t \upd t' \delta(X_{t'})$. More rigorously, it can defined by introducing a mollifier $\delta_\varepsilon$ of the Dirac distribution, e.g. $\delta_\varepsilon(X)=\varepsilon^{-1} {\bf 1}_{X\in [0,\varepsilon]}$, and taking $L_t=\lim_{\varepsilon\to 0^+}\int_0^t \upd t' \delta_\varepsilon(X_{t'})$. Intuitively, the local time in 0 represents the rescaled time the process spends in 0.

Incidentally, this proposition gives a way to define jumps and spikes precisely in the infinite $\gamma$ limit. A jump is simply a transition from $Q=0$ to $Q=1$ (resp. $1$ to $0$) while a spike is a transition from $Q=0$ to $Q=0$ (resp. $1$ to $1$) through some finite value of $Q$. Both types of transitions are instantaneous in real time $s$ but take a finite effective time $t$. This shows that a finer description is preserved by the effective time. The following proposition provides an example of such a finer quantity: the effective time itself.

\begin{corollary}
The effective time description is strictly finer than the real time description.
\end{corollary}
There exists anomalous quantities, i.e. quantities which can be computed in effective time but are hidden in the standard physical time description.
For example, the effective times to go up or to go down a spike of height $m$ are distributed with the same probability density $\mathscr{P}_m(t)$ of Laplace transform:
\begin{equation}
\tilde{\mathscr{P}}_m(\sigma)=\int_0^{+\infty} e^{-t\sigma} \mathscr{P}_m(t) \, \upd t =\frac{m\sqrt{2\sigma}}{\sinh m \sqrt{2\sigma}}
\end{equation}
This means that the effective time is an \emph{anomalous} observable in the sense that it is not determined entirely by the naive large $\gamma$ limit in real time $s$ which contains only discrete spikes and jumps. It has intrinsic fluctuations even when the sample of spikes of given height is fixed.

\paragraph{Proofs ---}
Let us start by looking precisely at what happens near the boundary $Q=0$ when $\gamma\rightarrow+\infty$ (the boundary $Q=1$ can be treated in the same way). When $Q$ is close to 0, equation \eqref{eq:thermal-t} becomes:
\begin{equation}
\upd Q_t= \frac{\lambda p}{\gamma Q_t^2} \,\upd t + \upd B_t,
\end{equation}
which reads in integral form:
\begin{equation}\label{eq:integrated}
Q_t=Q_0 + \int_0^{t}\frac{\lambda p}{\gamma Q_u^2} \upd u  + B_t.
\end{equation}
The integral is an increasing function of t which remains nearly constant on time intervals for which $Q_u\gg\gamma^{-1/2}$. Hence, when $\gamma\rightarrow+\infty$, this function only increases when $Q_u=0$. The Skorokhod lemma \cite{skorokhod1961,yen2013} is the key to understand precisely the large $\gamma$ limit.
\begin{lemma}[Skorokhod]
Let  $b(t)$, $t\in [0,+\infty[$ be a continuous function with $b(0)=0$, and let $x_0 \geq 0$. There is a unique pair of continuous functions $x(t)$, $l(t)$, for $t\in [0,+\infty[$, such that: 
\begin{itemize}
\item[(i)] $x(0)=x_0$, and $x(t)\geq 0$ for $t\in [0,+\infty[$,
\item[(ii)] $l(t)$ is non-decreasing,
\item[(iii)] $l(t)$ increases only at $x(t)=0$, i.e. $l(t)$ is constant on each interval
where $x(t)>0$,
\item[(iv)] $x(t)=x_0+l(t)+b(t)$ for $t\in [0,+\infty[$.
\end{itemize}
The solution is given by $l(t):=\max[0,-m(t)-x_0]$ and $x(t):=x_0 + b(t)+\max[0,-m(t)-x_0]$ where $m(t):=\min_{t'\leq t} b(t')$.
\end{lemma}
Equation \eqref{eq:integrated} is nearly a Skorokhod decomposition of the Brownian motion, i.e. with $b(t)=B_t$ and $x(t)=Q_t$. We get a true Skorokhod decomposition when $\gamma\rightarrow +\infty$ and the lemma guaranties the unicity of the solution. The fundamental trick is that the Skorokhod decomposition of the Brownian motion on $[0,+\infty[$ is known independently and can be found through Tanaka's formula \cite{tanaka1979,oksendal1992} for the Ito derivative of the absolute value of a Brownian motion :
\begin{equation}
|\tilde{B}_t| = \int_0^t \text{sgn}(\tilde{B}_u) \upd \tilde{B}_u + L_t,
\end{equation}
where $\tilde{B}_u$ is a Brownian motion and $L_t$ is the local time in 0 of this Brownian. But $B_t=\int_0^t \text{sgn}(\tilde{B}_u) \upd \tilde{B}_u$ is also a Brownian motion so we can write:
\begin{equation}
|\tilde{B}_t| = B_t + L_t,
\end{equation}
which, by unicity, is the infinite $\gamma$ limit of equation \eqref{eq:integrated} (up to the initial condition). We thus see that near $Q=0$, $Q$ behaves like the absolute value of a Brownian motion and that: 
\begin{equation}\label{eq:limitL}
\int_0^{t}\frac{\lambda p}{\gamma Q_u^2} \upd u \underset{\gamma \rightarrow +\infty}{\longrightarrow} L_t
\end{equation}
We can now get the connection between physical time and effective time. Near the boundary $Q=0$, $\upd t=\gamma Q_s^2 \upd s$. Inserting this change of variable in the l.h.s of \eqref{eq:limitL} removes the $\gamma$ and yields:
\begin{equation}\label{eq:local0}
s(t) = \frac{L_t}{\lambda p}
\end{equation}
We can apply the same reasoning near the boundary $Q=1$ to get that $Q_t$ is reflected by this boundary and that:
\begin{equation}
\int_0^{t}\frac{\lambda (1-p)}{\gamma(1- Q_u)^2} \upd u \underset{\gamma \rightarrow +\infty}{\longrightarrow} U_t
\end{equation}
where $U_t$ is the local time spent by $Q_t$ in $1$. Near this boundary the relation between physical and effective time can be found in the same way:
\begin{equation}
s(t) = \frac{U_t}{\lambda (1-p)}
\end{equation}
Out of the two boundaries, $L_t=U_t=0$, eq. \eqref{eq:thermal-t} shows that $Q_t$ is simply a Brownian motion and the physical time does not flow. Finally, we can put all the pieces together and we get that in the infinite $\gamma$ limit, $Q_t$ is a Brownian motion reflected in $0$ and $1$ or equivalently that $Q_t$ verifies \footnote{Such a decomposition could have also been obtained directly from a straightforward generalisation of Skorokhod's lemma in the strip $[0,1]$ instead of the open interval $[0,+\infty[$.}:
\begin{equation}
Q_t= Q_0 + B_t + L_t - U_t
\end{equation}
and the physical time is related to the effective time by:
\begin{equation}
s(t) = \frac{L_t}{\lambda p}+ \frac{U_t}{\lambda (1-p)}
\end{equation}
which is what we had claimed in the first proposition.

We may now prove the corollary using a standard result \cite{oksendal1992,feller2008}  for Brownian excursions. The time $t_1$ it takes for a Brownian motion starting from $0$ to reach a maximum $m$ and the time $t_2$ it then needs to go back to zero are independent random variables distributed with the same law of Laplace transform:
\begin{equation}
\tilde{\mathscr{P}}_m(\sigma)=\int_0^{+\infty} e^{-t\sigma} \mathscr{P}_m(t) \, \upd t =\frac{m\sqrt{2\sigma}}{\sinh m \sqrt{2\sigma}}
\end{equation}
Restricted to $m<1$, this is thus the probability distribution for the time it takes $Q_t$ to reach a maximum $m$ before eventually going back to $0$. Because the real time does not flow in the bulk, this excursion looks instantaneous when parametrised with $s$ --it is reduced to a spike-- and its finer structure, in this example the quadratic variation, is lost.

\paragraph{Applications ---}
The formalism previously introduced can be applied to describe the evolution of physical quantities when $\gamma \rightarrow +\infty$. A simple example one can consider is the linear entropy \footnote{Similar computations could be carried out with the Von Neumann entropy but the analysis would be mathematically much subtler because of the divergence of the logarithm in zero.} $S^L=1-\tr [\rho^2] = 2 \,Q(1-Q)$. In real time and for finite $\gamma$, It\^o's formula gives:
\begin{equation*}
\begin{split}
\upd S^L_s&= 2\lambda (1-2Q_s) (p-Q_s) \,\upd s \\
+& Q_s (1-Q_s) \left[2\sqrt{\gamma} (1-2Q_s)\upd W_s -2 \gamma Q_s(1-Q_s) \upd s\right]
\end{split}
\end{equation*}
The first term codes for the effect of the thermal bath and the second for the effect of the information extraction, the latter always decreasing the linear entropy on average. When $\gamma$ goes to infinity, the previous equation has no limit. Intuitively, in real time, $Q$ is almost surely equal to $0$ or $1$ and the linear entropy is thus almost always equal to zero, all the interesting fluctuations being lost. The latter can be recovered in effective time. Indeed, in effective time, when $\gamma\rightarrow\infty$, $\upd Q_t= \upd B_t + \upd L_t - \upd U_t$. Applying the It\^o formula to $S^L$ and noting that $\upd L$ (resp. $\upd U$) is only non zero when $Q=0$ (resp. $Q=1$) gives:
\begin{equation}
\upd S^L_t = 2(1-2Q_t)\, \upd B_t - 2\,\upd t + 2 (\upd L_t + \upd U_t)
\end{equation}
The effect of measurements appears clearly in the first two terms with a noise term and a deterministic negative drift. The effect of the bath is localised on the boundaries, i.e. on pure states, where it thus always increases the linear entropy. The details of this competition are simply lost when taking naively the $\gamma\rightarrow + \infty$ limit in real time.

The proposition also shows an interesting link with L\'evy processes. Near a boundary, say near $0$, the real time is given as a function of the effective time by eq. \eqref{eq:local0}. It is a standard result (see e.g. \cite{yen2013}) that $t(s)$, obtained by inverting
  the latter relation, is exactly the stable L\'evy process with index 1/2
and scale $\sqrt{2}\lambda p$. This means that, for $0 < s_1 < \cdots < s_n,$
\begin{eqnarray*} 
\expect \left[ e^{-\sigma_1 t(s_1)-\sigma_2 t((s_2)-t(s_1))-\cdots -\sigma_n
    (t(s_n)-t(s_{n-1}))}\right] & = & \\ 
& & \hspace{-6.5cm} e^{-s_1\lambda p\sqrt{2\sigma_1}-(s_2-s_1) \lambda p\sqrt{2\sigma_2}-\cdots-
  (s_n-s_{n-1})\lambda p\sqrt{2\sigma_n}}. 
\end{eqnarray*} 
In particular $\expect [e^{-\sigma t(s)}]=e^{-s\lambda p \sqrt{2\sigma}}$ whose
Laplace transform can be inverted to get that the probability density of
$t(s)$ is 
\begin{equation}
\upd P_s(t)=\frac{\upd t}{\sqrt{2\pi}}\frac{\lambda p s}{t^{3/2}}e^{-\frac{(\lambda p s)^2}{2t}}.
\end{equation}
Because of the presence of the second boundary, the description in terms of L\'evy processes is correct only near a boundary, i.e. for small jumps of $t$. In general the process $t$ is still infinitely divisible but not stable and its probability distribution has no simple closed form to our knowledge.
\paragraph{Conclusion ---}
We have argued that using a time proportional to the effect of measurements on the system provided a better parametrisation of the evolution in the limit of infinitely strong continuous measurements (infinite $\gamma$ limit). We have illustrated the benefits of this approach on the example of a continuously monitored qubit coupled to a thermal reservoir. Actually, our result is general in two dimensions and the dissipative coupling via a bath could have been replaced by an appropriately rescaled \footnote{If the dissipative coupling is replaced by a unitary evolution, the latter needs to be rescaled by $\sqrt{\gamma}$ to counter the Zeno effect and get meaningful results in the limit.} unitary evolution to yield the same process in the limit. In the infinite $\gamma$ limit, we have obtained a very simple description in terms of a reflected Brownian motion, unravelling a much finer structure than one would have gotten taking naively the limit in real time. In the limit, most quantities of interest can be computed using standard results on Brownian excursions.

Although our prescription for the time redefinition is very general, we have only treated in detail an example in two dimensions and with continuous measurements. We believe the same ideas could be used in the discrete case of iterated weak measurement and in higher dimensions but specific examples of interest are still to be worked out. Eventually, the method we propose is general enough that it could have applications to the analysis of other SDE's in the strong noise limit, e.g. in population dynamics and turbulence.

\begin{acknowledgments}
We thank Jean Bertoin for discussions. This work was supported in part by the Agence Nationale de la Recherche (ANR) contract ANR-14-CE25-0003-01.
\end{acknowledgments}

\bibliography{main}

\begin{thebibliography}{36}%
\makeatletter
\providecommand \@ifxundefined [1]{%
 \@ifx{#1\undefined}
}%
\providecommand \@ifnum [1]{%
 \ifnum #1\expandafter \@firstoftwo
 \else \expandafter \@secondoftwo
 \fi
}%
\providecommand \@ifx [1]{%
 \ifx #1\expandafter \@firstoftwo
 \else \expandafter \@secondoftwo
 \fi
}%
\providecommand \natexlab [1]{#1}%
\providecommand \enquote  [1]{``#1''}%
\providecommand \bibnamefont  [1]{#1}%
\providecommand \bibfnamefont [1]{#1}%
\providecommand \citenamefont [1]{#1}%
\providecommand \href@noop [0]{\@secondoftwo}%
\providecommand \href [0]{\begingroup \@sanitize@url \@href}%
\providecommand \@href[1]{\@@startlink{#1}\@@href}%
\providecommand \@@href[1]{\endgroup#1\@@endlink}%
\providecommand \@sanitize@url [0]{\catcode `\\12\catcode `\$12\catcode
  `\&12\catcode `\#12\catcode `\^12\catcode `\_12\catcode `\%12\relax}%
\providecommand \@@startlink[1]{}%
\providecommand \@@endlink[0]{}%
\providecommand \url  [0]{\begingroup\@sanitize@url \@url }%
\providecommand \@url [1]{\endgroup\@href {#1}{\urlprefix }}%
\providecommand \urlprefix  [0]{URL }%
\providecommand \Eprint [0]{\href }%
\providecommand \doibase [0]{http://dx.doi.org/}%
\providecommand \selectlanguage [0]{\@gobble}%
\providecommand \bibinfo  [0]{\@secondoftwo}%
\providecommand \bibfield  [0]{\@secondoftwo}%
\providecommand \translation [1]{[#1]}%
\providecommand \BibitemOpen [0]{}%
\providecommand \bibitemStop [0]{}%
\providecommand \bibitemNoStop [0]{.\EOS\space}%
\providecommand \EOS [0]{\spacefactor3000\relax}%
\providecommand \BibitemShut  [1]{\csname bibitem#1\endcsname}%
\let\auto@bib@innerbib\@empty
\bibitem [{\citenamefont {Wiseman}\ and\ \citenamefont
  {Milburn}(1993)}]{wiseman1993}%
  \BibitemOpen
  \bibfield  {author} {\bibinfo {author} {\bibfnamefont {H.~M.}\ \bibnamefont
  {Wiseman}}\ and\ \bibinfo {author} {\bibfnamefont {G.~J.}\ \bibnamefont
  {Milburn}},\ }\href {\doibase 10.1103/PhysRevLett.70.548} {\bibfield
  {journal} {\bibinfo  {journal} {Phys. Rev. Lett.}\ }\textbf {\bibinfo
  {volume} {70}},\ \bibinfo {pages} {548} (\bibinfo {year} {1993})}\BibitemShut
  {NoStop}%
\bibitem [{\citenamefont {Wiseman}(1994)}]{wiseman1994}%
  \BibitemOpen
  \bibfield  {author} {\bibinfo {author} {\bibfnamefont {H.~M.}\ \bibnamefont
  {Wiseman}},\ }\href {\doibase 10.1103/PhysRevA.49.2133} {\bibfield  {journal}
  {\bibinfo  {journal} {Phys. Rev. A}\ }\textbf {\bibinfo {volume} {49}},\
  \bibinfo {pages} {2133} (\bibinfo {year} {1994})}\BibitemShut {NoStop}%
\bibitem [{\citenamefont {Doherty}\ and\ \citenamefont
  {Jacobs}(1999)}]{doherty1999}%
  \BibitemOpen
  \bibfield  {author} {\bibinfo {author} {\bibfnamefont {A.~C.}\ \bibnamefont
  {Doherty}}\ and\ \bibinfo {author} {\bibfnamefont {K.}~\bibnamefont
  {Jacobs}},\ }\href@noop {} {\bibfield  {journal} {\bibinfo  {journal} {Phys.
  Rev. A}\ }\textbf {\bibinfo {volume} {60}},\ \bibinfo {pages} {2700}
  (\bibinfo {year} {1999})}\BibitemShut {NoStop}%
\bibitem [{\citenamefont {Jacobs}(2003)}]{jacobs2003}%
  \BibitemOpen
  \bibfield  {author} {\bibinfo {author} {\bibfnamefont {K.}~\bibnamefont
  {Jacobs}},\ }\href@noop {} {\bibfield  {journal} {\bibinfo  {journal}
  {Physical Review A}\ }\textbf {\bibinfo {volume} {67}},\ \bibinfo {pages}
  {030301} (\bibinfo {year} {2003})}\BibitemShut {NoStop}%
\bibitem [{\citenamefont {Combes}\ \emph {et~al.}(2008)\citenamefont {Combes},
  \citenamefont {Wiseman},\ and\ \citenamefont {Jacobs}}]{combes2008}%
  \BibitemOpen
  \bibfield  {author} {\bibinfo {author} {\bibfnamefont {J.}~\bibnamefont
  {Combes}}, \bibinfo {author} {\bibfnamefont {H.~M.}\ \bibnamefont {Wiseman}},
  \ and\ \bibinfo {author} {\bibfnamefont {K.}~\bibnamefont {Jacobs}},\
  }\href@noop {} {\bibfield  {journal} {\bibinfo  {journal} {Physical review
  letters}\ }\textbf {\bibinfo {volume} {100}},\ \bibinfo {pages} {160503}
  (\bibinfo {year} {2008})}\BibitemShut {NoStop}%
\bibitem [{\citenamefont {Combes}\ \emph {et~al.}(2015)\citenamefont {Combes},
  \citenamefont {Denney},\ and\ \citenamefont {Wiseman}}]{combes2015}%
  \BibitemOpen
  \bibfield  {author} {\bibinfo {author} {\bibfnamefont {J.}~\bibnamefont
  {Combes}}, \bibinfo {author} {\bibfnamefont {A.}~\bibnamefont {Denney}}, \
  and\ \bibinfo {author} {\bibfnamefont {H.~M.}\ \bibnamefont {Wiseman}},\
  }\href@noop {} {\bibfield  {journal} {\bibinfo  {journal} {Physical Review
  A}\ }\textbf {\bibinfo {volume} {91}},\ \bibinfo {pages} {022305} (\bibinfo
  {year} {2015})}\BibitemShut {NoStop}%
\bibitem [{\citenamefont {Bassi}\ and\ \citenamefont
  {Ghirardi}(2003)}]{bassi2003}%
  \BibitemOpen
  \bibfield  {author} {\bibinfo {author} {\bibfnamefont {A.}~\bibnamefont
  {Bassi}}\ and\ \bibinfo {author} {\bibfnamefont {G.}~\bibnamefont
  {Ghirardi}},\ }\href@noop {} {\bibfield  {journal} {\bibinfo  {journal}
  {Physics Reports}\ }\textbf {\bibinfo {volume} {379}},\ \bibinfo {pages}
  {257} (\bibinfo {year} {2003})}\BibitemShut {NoStop}%
\bibitem [{\citenamefont {Bassi}\ \emph {et~al.}(2013)\citenamefont {Bassi},
  \citenamefont {Lochan}, \citenamefont {Satin}, \citenamefont {Singh},\ and\
  \citenamefont {Ulbricht}}]{bassi2013}%
  \BibitemOpen
  \bibfield  {author} {\bibinfo {author} {\bibfnamefont {A.}~\bibnamefont
  {Bassi}}, \bibinfo {author} {\bibfnamefont {K.}~\bibnamefont {Lochan}},
  \bibinfo {author} {\bibfnamefont {S.}~\bibnamefont {Satin}}, \bibinfo
  {author} {\bibfnamefont {T.~P.}\ \bibnamefont {Singh}}, \ and\ \bibinfo
  {author} {\bibfnamefont {H.}~\bibnamefont {Ulbricht}},\ }\href@noop {}
  {\bibfield  {journal} {\bibinfo  {journal} {Reviews of Modern Physics}\
  }\textbf {\bibinfo {volume} {85}},\ \bibinfo {pages} {471} (\bibinfo {year}
  {2013})}\BibitemShut {NoStop}%
\bibitem [{\citenamefont {Guerlin}\ \emph {et~al.}(2007)\citenamefont
  {Guerlin}, \citenamefont {Bernu}, \citenamefont {Deleglise}, \citenamefont
  {Sayrin}, \citenamefont {Gleyzes}, \citenamefont {Kuhr}, \citenamefont
  {Brune}, \citenamefont {Raimond},\ and\ \citenamefont
  {Haroche}}]{guerlin2007}%
  \BibitemOpen
  \bibfield  {author} {\bibinfo {author} {\bibfnamefont {C.}~\bibnamefont
  {Guerlin}}, \bibinfo {author} {\bibfnamefont {J.}~\bibnamefont {Bernu}},
  \bibinfo {author} {\bibfnamefont {S.}~\bibnamefont {Deleglise}}, \bibinfo
  {author} {\bibfnamefont {C.}~\bibnamefont {Sayrin}}, \bibinfo {author}
  {\bibfnamefont {S.}~\bibnamefont {Gleyzes}}, \bibinfo {author} {\bibfnamefont
  {S.}~\bibnamefont {Kuhr}}, \bibinfo {author} {\bibfnamefont {M.}~\bibnamefont
  {Brune}}, \bibinfo {author} {\bibfnamefont {J.-M.}\ \bibnamefont {Raimond}},
  \ and\ \bibinfo {author} {\bibfnamefont {S.}~\bibnamefont {Haroche}},\
  }\href@noop {} {\bibfield  {journal} {\bibinfo  {journal} {Nature}\ }\textbf
  {\bibinfo {volume} {448}},\ \bibinfo {pages} {889} (\bibinfo {year}
  {2007})}\BibitemShut {NoStop}%
\bibitem [{\citenamefont {Gleyzes}\ \emph {et~al.}(2007)\citenamefont
  {Gleyzes}, \citenamefont {Kuhr}, \citenamefont {Guerlin}, \citenamefont
  {Bernu}, \citenamefont {Deleglise}, \citenamefont {Hoff}, \citenamefont
  {Brune}, \citenamefont {Raimond},\ and\ \citenamefont
  {Haroche}}]{gleyzes2007}%
  \BibitemOpen
  \bibfield  {author} {\bibinfo {author} {\bibfnamefont {S.}~\bibnamefont
  {Gleyzes}}, \bibinfo {author} {\bibfnamefont {S.}~\bibnamefont {Kuhr}},
  \bibinfo {author} {\bibfnamefont {C.}~\bibnamefont {Guerlin}}, \bibinfo
  {author} {\bibfnamefont {J.}~\bibnamefont {Bernu}}, \bibinfo {author}
  {\bibfnamefont {S.}~\bibnamefont {Deleglise}}, \bibinfo {author}
  {\bibfnamefont {U.~B.}\ \bibnamefont {Hoff}}, \bibinfo {author}
  {\bibfnamefont {M.}~\bibnamefont {Brune}}, \bibinfo {author} {\bibfnamefont
  {J.-M.}\ \bibnamefont {Raimond}}, \ and\ \bibinfo {author} {\bibfnamefont
  {S.}~\bibnamefont {Haroche}},\ }\href@noop {} {\bibfield  {journal} {\bibinfo
   {journal} {Nature}\ }\textbf {\bibinfo {volume} {446}},\ \bibinfo {pages}
  {297} (\bibinfo {year} {2007})}\BibitemShut {NoStop}%
\bibitem [{\citenamefont {Murch}\ \emph {et~al.}(2013)\citenamefont {Murch},
  \citenamefont {Weber}, \citenamefont {Macklin},\ and\ \citenamefont
  {Siddiqi}}]{murch2013}%
  \BibitemOpen
  \bibfield  {author} {\bibinfo {author} {\bibfnamefont {K.}~\bibnamefont
  {Murch}}, \bibinfo {author} {\bibfnamefont {S.}~\bibnamefont {Weber}},
  \bibinfo {author} {\bibfnamefont {C.}~\bibnamefont {Macklin}}, \ and\
  \bibinfo {author} {\bibfnamefont {I.}~\bibnamefont {Siddiqi}},\ }\href@noop
  {} {\bibfield  {journal} {\bibinfo  {journal} {Nature}\ }\textbf {\bibinfo
  {volume} {502}},\ \bibinfo {pages} {211} (\bibinfo {year}
  {2013})}\BibitemShut {NoStop}%
\bibitem [{\citenamefont {Weber}\ \emph {et~al.}(2014)\citenamefont {Weber},
  \citenamefont {Chantasri}, \citenamefont {Dressel}, \citenamefont {Jordan},
  \citenamefont {Murch},\ and\ \citenamefont {Siddiqi}}]{weber2014}%
  \BibitemOpen
  \bibfield  {author} {\bibinfo {author} {\bibfnamefont {S.}~\bibnamefont
  {Weber}}, \bibinfo {author} {\bibfnamefont {A.}~\bibnamefont {Chantasri}},
  \bibinfo {author} {\bibfnamefont {J.}~\bibnamefont {Dressel}}, \bibinfo
  {author} {\bibfnamefont {A.}~\bibnamefont {Jordan}}, \bibinfo {author}
  {\bibfnamefont {K.}~\bibnamefont {Murch}}, \ and\ \bibinfo {author}
  {\bibfnamefont {I.}~\bibnamefont {Siddiqi}},\ }\href@noop {} {\bibfield
  {journal} {\bibinfo  {journal} {Nature}\ }\textbf {\bibinfo {volume} {511}},\
  \bibinfo {pages} {570} (\bibinfo {year} {2014})}\BibitemShut {NoStop}%
\bibitem [{\citenamefont {Bauer}\ \emph {et~al.}(2015)\citenamefont {Bauer},
  \citenamefont {Bernard},\ and\ \citenamefont {Tilloy}}]{jumps}%
  \BibitemOpen
  \bibfield  {author} {\bibinfo {author} {\bibfnamefont {M.}~\bibnamefont
  {Bauer}}, \bibinfo {author} {\bibfnamefont {D.}~\bibnamefont {Bernard}}, \
  and\ \bibinfo {author} {\bibfnamefont {A.}~\bibnamefont {Tilloy}},\ }\href
  {http://stacks.iop.org/1751-8121/48/i=25/a=25FT02} {\bibfield  {journal}
  {\bibinfo  {journal} {Journal of Physics A: Mathematical and Theoretical}\
  }\textbf {\bibinfo {volume} {48}},\ \bibinfo {pages} {25FT02} (\bibinfo
  {year} {2015})}\BibitemShut {NoStop}%
\bibitem [{\citenamefont {Nagourney}\ \emph {et~al.}(1986)\citenamefont
  {Nagourney}, \citenamefont {Sandberg},\ and\ \citenamefont
  {Dehmelt}}]{nagourney}%
  \BibitemOpen
  \bibfield  {author} {\bibinfo {author} {\bibfnamefont {W.}~\bibnamefont
  {Nagourney}}, \bibinfo {author} {\bibfnamefont {J.}~\bibnamefont {Sandberg}},
  \ and\ \bibinfo {author} {\bibfnamefont {H.}~\bibnamefont {Dehmelt}},\ }\href
  {\doibase 10.1103/PhysRevLett.56.2797} {\bibfield  {journal} {\bibinfo
  {journal} {Phys. Rev. Lett.}\ }\textbf {\bibinfo {volume} {56}},\ \bibinfo
  {pages} {2797} (\bibinfo {year} {1986})}\BibitemShut {NoStop}%
\bibitem [{\citenamefont {Bergquist}\ \emph {et~al.}(1986)\citenamefont
  {Bergquist}, \citenamefont {Hulet}, \citenamefont {Itano},\ and\
  \citenamefont {Wineland}}]{bergquist}%
  \BibitemOpen
  \bibfield  {author} {\bibinfo {author} {\bibfnamefont {J.~C.}\ \bibnamefont
  {Bergquist}}, \bibinfo {author} {\bibfnamefont {R.~G.}\ \bibnamefont
  {Hulet}}, \bibinfo {author} {\bibfnamefont {W.~M.}\ \bibnamefont {Itano}}, \
  and\ \bibinfo {author} {\bibfnamefont {D.~J.}\ \bibnamefont {Wineland}},\
  }\href {\doibase 10.1103/PhysRevLett.57.1699} {\bibfield  {journal} {\bibinfo
   {journal} {Phys. Rev. Lett.}\ }\textbf {\bibinfo {volume} {57}},\ \bibinfo
  {pages} {1699} (\bibinfo {year} {1986})}\BibitemShut {NoStop}%
\bibitem [{\citenamefont {Sauter}\ \emph {et~al.}(1986)\citenamefont {Sauter},
  \citenamefont {Neuhauser}, \citenamefont {Blatt},\ and\ \citenamefont
  {Toschek}}]{sauter}%
  \BibitemOpen
  \bibfield  {author} {\bibinfo {author} {\bibfnamefont {T.}~\bibnamefont
  {Sauter}}, \bibinfo {author} {\bibfnamefont {W.}~\bibnamefont {Neuhauser}},
  \bibinfo {author} {\bibfnamefont {R.}~\bibnamefont {Blatt}}, \ and\ \bibinfo
  {author} {\bibfnamefont {P.}~\bibnamefont {Toschek}},\ }\href@noop {}
  {\bibfield  {journal} {\bibinfo  {journal} {Phys. Rev. Lett.}\ }\textbf
  {\bibinfo {volume} {57}},\ \bibinfo {pages} {1696} (\bibinfo {year}
  {1986})}\BibitemShut {NoStop}%
\bibitem [{\citenamefont {Tilloy}\ \emph {et~al.}(2015)\citenamefont {Tilloy},
  \citenamefont {Bauer},\ and\ \citenamefont {Bernard}}]{spikes}%
  \BibitemOpen
  \bibfield  {author} {\bibinfo {author} {\bibfnamefont {A.}~\bibnamefont
  {Tilloy}}, \bibinfo {author} {\bibfnamefont {M.}~\bibnamefont {Bauer}}, \
  and\ \bibinfo {author} {\bibfnamefont {D.}~\bibnamefont {Bernard}},\ }\href
  {\doibase 10.1103/PhysRevA.92.052111} {\bibfield  {journal} {\bibinfo
  {journal} {Phys. Rev. A}\ }\textbf {\bibinfo {volume} {92}},\ \bibinfo
  {pages} {052111} (\bibinfo {year} {2015})}\BibitemShut {NoStop}%
\bibitem [{Note1()}]{Note1}%
  \BibitemOpen
  \bibinfo {note} {Other prescriptions with the same quadratic scaling are
  possible, e.g. $\Delta t_n= \protect \text {Tr}\protect \tmspace +\thinmuskip
  {.1667em}[ (D(\rho _{n})-D(\rho _{n-1}))^2]$ where $D(\cdot )$ denotes the
  diagonal part of a matrix in the measurement pointer basis.}\BibitemShut
  {Stop}%
\bibitem [{\citenamefont {Barchielli}(1986)}]{barchielli1986}%
  \BibitemOpen
  \bibfield  {author} {\bibinfo {author} {\bibfnamefont {A.}~\bibnamefont
  {Barchielli}},\ }\href {\doibase 10.1103/PhysRevA.34.1642} {\bibfield
  {journal} {\bibinfo  {journal} {Phys. Rev. A}\ }\textbf {\bibinfo {volume}
  {34}},\ \bibinfo {pages} {1642} (\bibinfo {year} {1986})}\BibitemShut
  {NoStop}%
\bibitem [{\citenamefont {Belavkin}(1992)}]{belavkin1992}%
  \BibitemOpen
  \bibfield  {author} {\bibinfo {author} {\bibfnamefont {V.-P.}\ \bibnamefont
  {Belavkin}},\ }\href@noop {} {\bibfield  {journal} {\bibinfo  {journal}
  {Comm. Math. Phys.}\ }\textbf {\bibinfo {volume} {146}},\ \bibinfo {pages}
  {611} (\bibinfo {year} {1992})}\BibitemShut {NoStop}%
\bibitem [{\citenamefont {Jacobs}\ and\ \citenamefont
  {Steck}(2006)}]{jacobs2006}%
  \BibitemOpen
  \bibfield  {author} {\bibinfo {author} {\bibfnamefont {K.}~\bibnamefont
  {Jacobs}}\ and\ \bibinfo {author} {\bibfnamefont {D.~A.}\ \bibnamefont
  {Steck}},\ }\href@noop {} {\bibfield  {journal} {\bibinfo  {journal}
  {Contemporary Physics}\ }\textbf {\bibinfo {volume} {47}},\ \bibinfo {pages}
  {279} (\bibinfo {year} {2006})}\BibitemShut {NoStop}%
\bibitem [{\citenamefont {Wiseman}\ and\ \citenamefont
  {Milburn}(2009)}]{wiseman2009}%
  \BibitemOpen
  \bibfield  {author} {\bibinfo {author} {\bibfnamefont {H.~M.}\ \bibnamefont
  {Wiseman}}\ and\ \bibinfo {author} {\bibfnamefont {G.~J.}\ \bibnamefont
  {Milburn}},\ }\href@noop {} {\emph {\bibinfo {title} {Quantum measurement and
  control}}}\ (\bibinfo  {publisher} {Cambridge University Press},\ \bibinfo
  {year} {2009})\BibitemShut {NoStop}%
\bibitem [{\citenamefont {Bauer}\ and\ \citenamefont {Bernard}(2014)}]{lmp}%
  \BibitemOpen
  \bibfield  {author} {\bibinfo {author} {\bibfnamefont {M.}~\bibnamefont
  {Bauer}}\ and\ \bibinfo {author} {\bibfnamefont {D.}~\bibnamefont
  {Bernard}},\ }\href@noop {} {\bibfield  {journal} {\bibinfo  {journal}
  {Letters in Mathematical Physics}\ }\textbf {\bibinfo {volume} {104}},\
  \bibinfo {pages} {707} (\bibinfo {year} {2014})}\BibitemShut {NoStop}%
\bibitem [{Note2()}]{Note2}%
  \BibitemOpen
  \bibinfo {note} {One can give a heuristic argument for the existence of
  spikes. Near a boundary say $Q=0$, $\protect \mathrm {d}Q_s\simeq \lambda p
  \protect \mathrm {d}s +\protect \sqrt {\gamma }Q_s \protect \mathrm {d}W_s$.
  One can show that for large $\gamma $, this means that the distribution of
  $Q$ is given by $\protect \mathds {P}[Q<q]=e^{-2/(\gamma \lambda p q)}$ and
  that $Q_{s+1/\gamma }$ is weakly correlated with $Q_s$. As a result, the
  maximum $M_{a,b}$ of $Q$ in an interval $[a,b]$ has a distribution which can
  be approximated by $\protect \mathds {P}[M_{a,b}<m]\simeq \left
  (e^{-2/(\gamma \lambda p m)}\right )^{\gamma (b-a)}$. The $\gamma $ cancel
  out which means that excursions with height of order $0$ (but vanishing
  width) persist in the limit: these are the spikes.}\BibitemShut {Stop}%
\bibitem [{Note3()}]{Note3}%
  \BibitemOpen
  \bibinfo {note} {We borrow this terminology from quantum or statistical field
  theory and from fluid turbulence. For instance, observables linked to
  dissipative processes in Burgers' turbulence are localised on velocity shocks
  \cite {polyakov1993,polyakov1995,bernard1998}. In a way similar to Burgers'
  turbulence, we can consider ``anomalous observables'' localised on spikes and
  jumps.}\BibitemShut {Stop}%
\bibitem [{\citenamefont {Skorokhod}(1961)}]{skorokhod1961}%
  \BibitemOpen
  \bibfield  {author} {\bibinfo {author} {\bibfnamefont {A.~V.}\ \bibnamefont
  {Skorokhod}},\ }\href@noop {} {\bibfield  {journal} {\bibinfo  {journal}
  {Theory of Probability \& Its Applications}\ }\textbf {\bibinfo {volume}
  {6}},\ \bibinfo {pages} {264} (\bibinfo {year} {1961})}\BibitemShut {NoStop}%
\bibitem [{\citenamefont {Yen}\ and\ \citenamefont {Yor}(2013)}]{yen2013}%
  \BibitemOpen
  \bibfield  {author} {\bibinfo {author} {\bibfnamefont {J.-Y.}\ \bibnamefont
  {Yen}}\ and\ \bibinfo {author} {\bibfnamefont {M.}~\bibnamefont {Yor}},\ }in\
  \href {\doibase 10.1007/978-3-319-01270-4_5} {\emph {\bibinfo {booktitle}
  {Local Times and Excursion Theory for Brownian Motion}}},\ \bibinfo {series}
  {Lecture Notes in Mathematics}, Vol.\ \bibinfo {volume} {2088}\ (\bibinfo
  {publisher} {Springer International Publishing},\ \bibinfo {year}
  {2013})\BibitemShut {NoStop}%
\bibitem [{\citenamefont {Tanaka}\ \emph {et~al.}(1979)\citenamefont {Tanaka}
  \emph {et~al.}}]{tanaka1979}%
  \BibitemOpen
  \bibfield  {author} {\bibinfo {author} {\bibfnamefont {H.}~\bibnamefont
  {Tanaka}} \emph {et~al.},\ }\href@noop {} {\bibfield  {journal} {\bibinfo
  {journal} {Hiroshima Math. J}\ }\textbf {\bibinfo {volume} {9}},\ \bibinfo
  {pages} {163} (\bibinfo {year} {1979})}\BibitemShut {NoStop}%
\bibitem [{\citenamefont {Oksendal}(1992)}]{oksendal1992}%
  \BibitemOpen
  \bibfield  {author} {\bibinfo {author} {\bibfnamefont {B.}~\bibnamefont
  {Oksendal}},\ }\href@noop {} {\emph {\bibinfo {title} {Stochastic
  differential equations: an introduction with applications}}},\ Vol.~\bibinfo
  {volume} {5}\ (\bibinfo  {publisher} {Springer New York},\ \bibinfo {year}
  {1992})\BibitemShut {NoStop}%
\bibitem [{Note4()}]{Note4}%
  \BibitemOpen
  \bibinfo {note} {Such a decomposition could have also been obtained directly
  from a straightforward generalisation of Skorokhod's lemma in the strip
  $[0,1]$ instead of the open interval $[0,+\infty [$.}\BibitemShut {Stop}%
\bibitem [{\citenamefont {Feller}(2008)}]{feller2008}%
  \BibitemOpen
  \bibfield  {author} {\bibinfo {author} {\bibfnamefont {W.}~\bibnamefont
  {Feller}},\ }\href@noop {} {\emph {\bibinfo {title} {An introduction to
  probability theory and its applications}}},\ Vol.~\bibinfo {volume} {2}\
  (\bibinfo  {publisher} {John Wiley \& Sons},\ \bibinfo {year}
  {2008})\BibitemShut {NoStop}%
\bibitem [{Note5()}]{Note5}%
  \BibitemOpen
  \bibinfo {note} {Similar computations could be carried out with the Von
  Neumann entropy but the analysis would be mathematically much subtler because
  of the divergence of the logarithm in zero.}\BibitemShut {Stop}%
\bibitem [{Note6()}]{Note6}%
  \BibitemOpen
  \bibinfo {note} {If the dissipative coupling is replaced by a unitary
  evolution, the latter needs to be rescaled by $\protect \sqrt {\gamma }$ to
  counter the Zeno effect and get meaningful results in the limit.}\BibitemShut
  {Stop}%
\bibitem [{\citenamefont {Polyakov}(1993)}]{polyakov1993}%
  \BibitemOpen
  \bibfield  {author} {\bibinfo {author} {\bibfnamefont {A.~M.}\ \bibnamefont
  {Polyakov}},\ }\href@noop {} {\bibfield  {journal} {\bibinfo  {journal}
  {Nuclear Physics B}\ }\textbf {\bibinfo {volume} {396}},\ \bibinfo {pages}
  {367} (\bibinfo {year} {1993})}\BibitemShut {NoStop}%
\bibitem [{\citenamefont {Polyakov}(1995)}]{polyakov1995}%
  \BibitemOpen
  \bibfield  {author} {\bibinfo {author} {\bibfnamefont {A.~M.}\ \bibnamefont
  {Polyakov}},\ }\href@noop {} {\bibfield  {journal} {\bibinfo  {journal}
  {Physical Review E}\ }\textbf {\bibinfo {volume} {52}},\ \bibinfo {pages}
  {6183} (\bibinfo {year} {1995})}\BibitemShut {NoStop}%
\bibitem [{\citenamefont {Bernard}\ and\ \citenamefont
  {Gawedzki}(1998)}]{bernard1998}%
  \BibitemOpen
  \bibfield  {author} {\bibinfo {author} {\bibfnamefont {D.}~\bibnamefont
  {Bernard}}\ and\ \bibinfo {author} {\bibfnamefont {K.}~\bibnamefont
  {Gawedzki}},\ }\href@noop {} {\bibfield  {journal} {\bibinfo  {journal}
  {Journal of Physics A: Mathematical and General}\ }\textbf {\bibinfo {volume}
  {31}},\ \bibinfo {pages} {8735} (\bibinfo {year} {1998})}\BibitemShut
  {NoStop}%
\end{thebibliography}%

\end{document}